\documentclass[twocolumn]{article}
\usepackage{tda}

\title{Challenges in Reconstructing Shapes from\\Euler Characteristic Curves}
\author{Brittany Terese Fasy\thanks{Depart. of Mathematical Sciences,
    Montana State U.} \thanks{School of Computing,
            Montana State
            U. \newline
   {\scriptsize {\tt \{brittany.fasy, david.millman, annaschenfisch\}@montana.edu}
    {\tt \{samuel.micka, lucia.williams\}@msu.montana.edu}
    }}
        \and
        Samuel Micka\footnotemark[2]
        \and
        David L. Millman\footnotemark[2]
        \and
        Anna Schenfisch\footnotemark[1]
        \and
        Lucia Williams\footnotemark[2]
    }

\begin{document}
\thispagestyle{empty}
\maketitle
\section*{Abstract}

Shape recognition and classification is a problem with a wide variety of
applications. Several recent works have demonstrated that topological
descriptors can be used as summaries of shapes and utilized to compute
distances. In this abstract, we explore the use of a finite number of Euler
Characteristic Curves (ECC) to reconstruct plane graphs.
We highlight difficulties that occur when attempting to adopt approaches for
reconstruction with persistence diagrams to reconstruction with ECCs.
 Furthermore, we highlight specific arrangements of
vertices that create problems for reconstruction and present several
observations about how they affect the  ECC-based reconstruction. Finally, we show that plane graphs
without degree two vertices can be reconstructed using a finite number of ECCs.

\section{Introduction}
\label{sec:intro}

Shape comparison and classification is a common task in the field of computer
science, with applications in graphics, geometry, machine learning, and several
other research fields.  The problem has been well-studied in $\R^3$, with several
approaches described in the survey \cite{Tangelder2007}.  One relatively new
approach to the problem involves utilizing topological descriptors to represent
and compare the shapes.  In \cite{turner2014persistent}, Turner et al.\  proposed
the use of the zero- and one-dimensional persistence diagrams from lower-star
filtrations to compare triangulations of $S^k$ in $\R^d$, for $d > k$.
We call the mapping of a shape
to to a parameterized set
of diagrams the \emph{persistent homology transform} (PHT).
Their
main result (Cor.\ $3.4$ of \cite{turner2014persistent})
showed that the persistent homology transform~(PHT) is injective for comparing
triangulations of $\sph^2$ or $\sph^1$ embedded in~$\R^3$
(or triangulations of $\sph^1$ in $\R^2$), and thus
can be used to distinguish different shapes.
Turner et al.\ also extend the idea of the PHT to the
Euler Characteristic Curve (ECC) and describe the Euler Characteristic Transform (ECT),
a topological summary that records changes in the Euler Characteristic across a height
parameter, again from all directions.  Finally, using experimental results, the authors show
that the PHT and ECT performed well in clustering tasks.  In \cite{crawford2016functional},
Crawford et al. extend this work by proposing the smooth Euler Characteristic
Transform (SECT), a functional variant of the ECT with favorable properties for
analysis. They show that features derived from the SECT of tumor shapes are
better predictors of clinical outcomes of patients than other traditional~features.

The proof of injectivity (i.e., that a shape can be reconstructed
from the PHT or the ECC) uses an infinite set of a directions; however, using an
infinite set of directions is infeasible for computational purposes.  Thus,
both~\cite{crawford2016functional,turner2014persistent}
use sampling a finite set of directions for the
height filtrations in order to apply the technique to shape comparison.
In \cite{belton2018learning}, Belton et al.
present an algorithm for reconstructing plane graphs using a quadratic (hence, finite) number of persistence diagrams.
Simultaneous to that result, other researchers also attempted to give a finite number of
directions sufficient to fully determine a shape. Both
\cite{curry2018directions} and \cite{ghrist2018euler} give upper bounds on the
number of directions needed to determine a hidden shape in~$\R ^d$.  In order to
do this, they make assumptions about the curvature and geometry of the input
shape.  In our work, by contrast, we restrict to plane graphs, but
make no restrictions on curvature.

Here, we attempt to extend the work of \cite{belton2018learning} on the PHT to the ECT.
However,  difficulties arise
when using ECCs because they do not encode information about every vertex from
every direction, as a persistence diagram does when on-diagonal points are
included.  We show that, while the number of directions needed to give an ECT unique
to the input graph is linear in the number of vertices of the graph, it is difficult to determine which directions generate the necessary ECCs. As we will see, the main difficulty lies with the presence of degree two vertices.

\section{Background}
\label{sec:background}
In this paper, we focus on a subset of finite simplicial complexes
that are composed of only edges and vertices and are provided with a planar
straight-line
embedding in~$\R ^2$. We refer to these simplicial complexes as
\emph{plane~graphs}.  We refer the reader
to~\cite{edelsbrunner2010computational} for a general background on
persistent homology, and only present the necessary content here.

\paragraph{Assumptions} Let $K$ be a plane graph.  In what follows, we
assume that the vertices of $K$ have distinct $x$- and $y$-coordinates from one another. Furthermore, we assume that no three vertices are collinear.

\paragraph{Lower-Star Filtration} Let $\sph ^1$ be the unit sphere in $\R^2$.
Consider $\dir \in \sph^1$, i.e., a direction vector in~$\R^2$; we define the
\emph{lower-star filtration} with respect to $\dir$.  Let $\hFiltFun{\dir}:
\simComp\rightarrow \R$ be defined for a simplex $\sigma \subseteq \simComp$ by
$\hFiltFun{\dir}(\sigma) = \max_{v \in \sigma} \dprod{v}{\dir},$
where~$\dprod{x}{y}$ is the inner (dot) product and measures height in the
direction of unit vector $y$.  Intuitively, the height of $\sigma$ with respect
to~$\dir$ is the
maximum ``height'' of all vertices in~$\sigma$.  Then, for each $h \in
\R$, the
subcomplex~$\simComp_h:=\hFiltFunT{\dir}{(-\infty,h]}$ is composed of all
simplices that lie entirely below or at the height~$h$, with respect to the
direction $\dir$.  The lower-star filtration is sequence of subcomplexes
$\simComp_h$, where $h$ increases from $-\infty$ to $\infty$; notice that
$\simComp_h$ only changes when $h$ is the height of a vertex of $K$.

When we observe a
difference between $\simComp_{h-\epsilon}$ and~$\simComp_{h+\epsilon}$, we know that we have encountered a vertex. As in \cite{belton2018learning}, we define a structure to
encode what we know about this vertex in $\R^2$.
Given $\dir \in \sph^1$, and a height $h \in \R$, the
\emph{filtration line at height $h$} is the line, denoted $\pLine{\dir}{h}$,
 perpendicular to direction $\dir$ and at height~$h$ in direction $\dir$.
Given a finite set of vertices $V \subset \R^2$, the
\emph{filtration lines of~$V$} are the set of lines
$$
    \pdpLines{\dir}{V} = \{ \pLine{\dir}{h}~|~\exists v \in V
    ~\text{s.t.}~ h = v\cdot s \}.
$$
 Further, $\pdpLines{\dir}{V}$ will contain $|V|$ lines
if and only if no two vertices have the same height in direction
$\dir$.  Our assumptions guarantee distinct vertex heights
only for $(0,1)$, $(0,-1)$, $(1,0)$, and $(-1,0)$, referred to as the
\emph{cardinal directions}.
In~\cite{belton2018learning}, every line in $\pdpLines{\dir}{V}$ can be read off
of the persistence diagram, as every simplex corresponds to either a birth or
the death of a homology class. Next, we observe that we cannot witness all such
lines for another topological descriptor, the Euler Characteristic Curve.

\paragraph{Euler Characteristic Curves}
The Euler characteristic of a plane graph $K=(V,E)$ is $|V| - |E|$.
The \emph{Euler Characteristic Curve} (ECC) is the piecewise step
function of the Euler characteristic, whose domain is subcomplexes of
a filtration defined by some parameterization of $\simComp$.
In this paper, the parameter is the \emph{height} of a lower-star filtration.
Specifically, we define $\ecc{\dir}{\simComp} \colon \R \to \Z$
to be the function that maps a height $h$ to the Euler Characteristic of
$\simComp_{h}$.  Every
 change in the ECC corresponds to a filtration line from that
direction, but not vice versa.
For example, if an edge and vertex appear at the same height,
then the ECC does not~change.
We now refine our definition of filtration lines:
\begin{align*}
    \pLines{\dir}{V} = \{ \pLine{\dir}{h}&~|~\exists \epsilon_0 >0
    ~\text{s.t.}~ \forall \epsilon \in (0, \epsilon_0), \\
   & \ecc{\dir}{\simComp}(h-\epsilon) \neq \ecc{\dir}{\simComp}(h+\epsilon) \}.
\end{align*}
This set corresponds to the subset of vertices in $V$ that are \emph{witnessed} from
$\dir$ through the ECC $\ecc{\dir}{\simComp}$.
As such, we refer to these lines as \emph{witnessed lines}.
We note that the only time that a vertex is not witnessed is if the vertex is included
in the filtration at the same time as an edge because the vertex being added will
be cancelled out by the inclusion of the edge.
Furthermore, we note that $v$ lying on a filtration line from $\dir$
does not necessarily imply that
$v$ is witnessed from $\dir$, i.e., it could lie on a witness line for another vertex
if they lie at the same height from $\dir$.

\section{Towards Vertex Reconstruction}
\label{sec:degTwoScary}
We are interested in reconstructing a plane graph from ECCs from a finite number
of directions.  While three directions was sufficient for reconstructing
vertices using
persistence diagrams, ECCs contain strictly less information in each direction.
We observe the existence of a
linear number of directions that allows to fully reconstruct
the vertices of a plane~graph:
\begin{proposition}[ECC Existence]
    Given a plane graph $K=(V,E)$ with $|V| = n$, there exist $3n$ directions
    that can be used to reconstruct all vertices in~$V$.
\label{prop:nDirExist}
\end{proposition}
The proof of this claim may be found
in \appendref{proofs}. We note that while $3n$ directions are sufficient, this bound
is likely not tight.

Initially, attempting to use the techniques in
\cite{belton2018learning} seems promising for plane graph
reconstruction using ECCs, i.e., we can define a correspondence
between three-way witness line intersections (from carefully
chosen directions) and vertices.
However, certain types of vertices introduce difficulties.
\begin{figure}[]
  \centering
  \includegraphics[width=.3\textwidth]{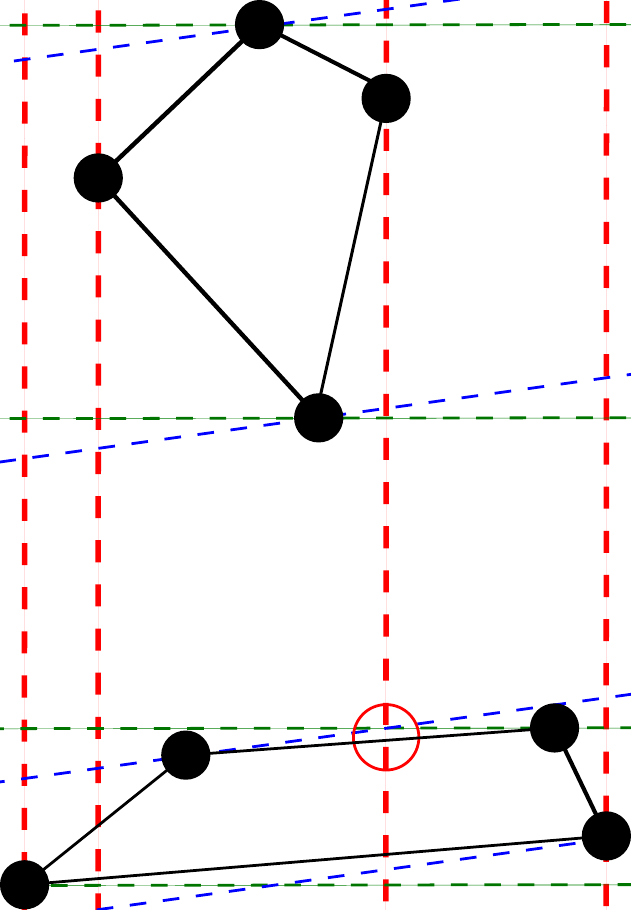}
  \caption{Scenario where a degree two vertex not witnessed
  by the cardinal directions can create a three-way line intersection
  where a vertex does not exist. The three-way intersection
  without a vertex is circled in red.}
  \label{fig:deg2Trick}
\end{figure}
For example, consider \figref{deg2Trick}. A degree two vertex is not witnessed
by any of the witness lines from the cardinal directions $(1,0), (0,1), (-1,0)$
and $(0,-1)$. However, we would like to generate a correspondence between
three-way intersections of witness lines and non degree two vertices. If we use
the technique described in Theorem $5$ of \cite{belton2018learning} to choose
such a direction, that direction creates a witness line that causes
a three-way intersection not corresponding to a vertex.  In fact, when degree
two vertices are introduced to the plane graph, several problems arise.  We
discuss these problems in detail in \secref{degTwoProb}.

\section{Degree Two Challenges}
\label{sec:degTwoProb}
Degree two vertices introduce several complications in finding witness
directions, because
degree two vertices can have an arbitrarily small
region on $\sph^1$ from which they can be witnessed.
\begin{figure}[]
  \centering
  \includegraphics[height=1.5in]{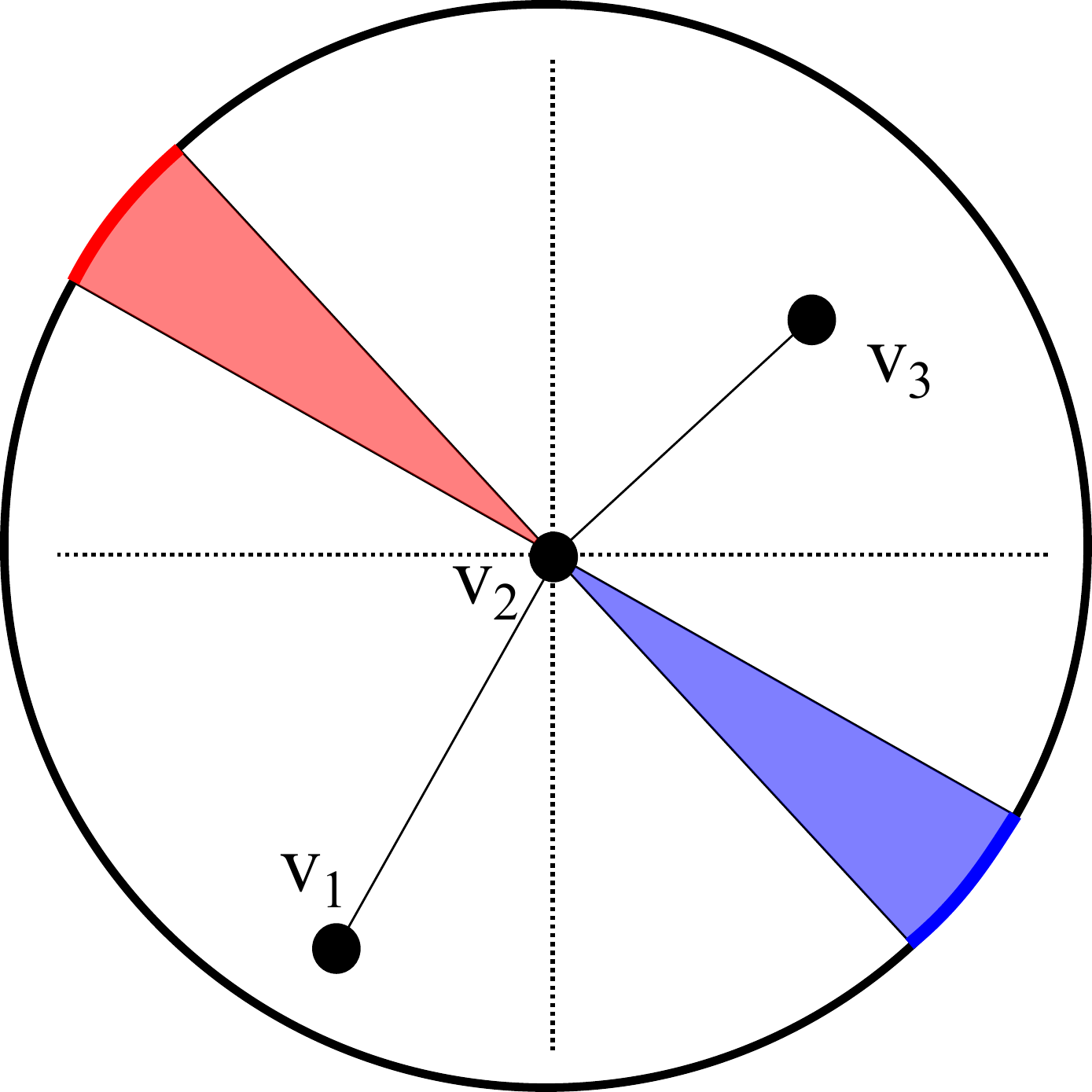}
  \caption{Case where $v_1$, $v_2$, and $v_3$ are nearly collinear.
As the vertices approach collinear the region on $\sph^1$ containing directions
which will witness $v_2$ grows arbitrarily small.}
  \label{fig:deg2Colin}
\end{figure}
For example, in~\figref{deg2Colin}~the vertices $v_1$, $v_2$, and
$v_3$ are nearly collinear.
In order to witness $v_2$, we must choose directions from within
the red region, where a decrease in the ECC will be observed,
or from the blue region, where an increase in the ECC will be observed.
However, these these regions becomes arbitrarily
small as $v_1$, $v_2$ and~$v_3$ approach collinear.

As mentioned earlier, degree two vertices
can also introduce additional
ambiguities when witnessing non-degree two vertices.
Recall the example found in \figref{deg2Trick}
and the discussion in \secref{degTwoScary}.

Despite these difficulties, several situations exist in which
degree two vertices can be witnessed.
The following propositions summarize these scenarios. Proofs
are provided in \appendref{proofs}.
For clarity, we discuss quadrants as though $v_2$ is located at
the origin. However, note that the following propositions also apply
to arrangements with similar orientations and angles.
\begin{figure}[]
  \centering
  \includegraphics[height=1.5in]{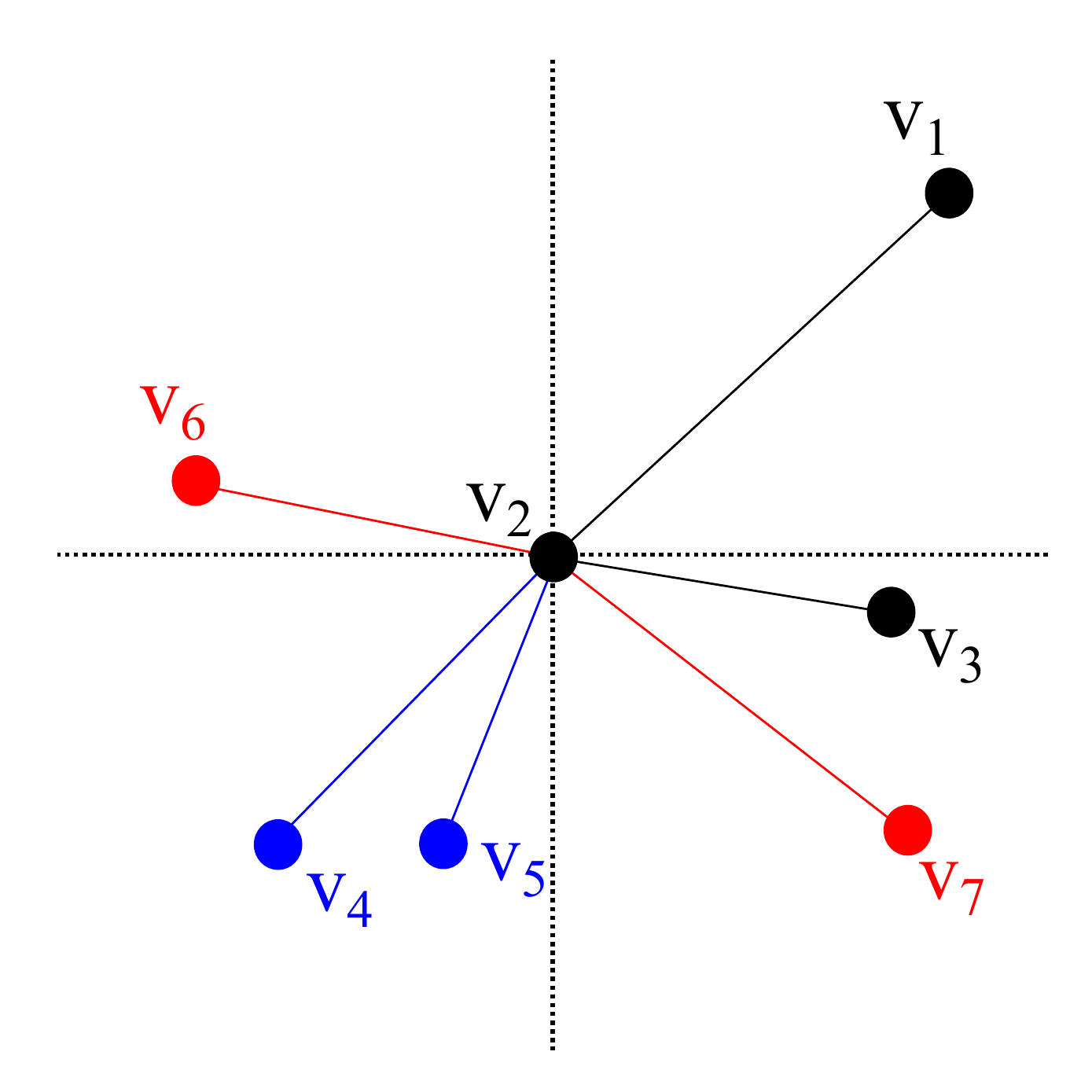}
  \caption{Different scenarios of edge embeddings for
  degree two vertices. We consider $v_2$ a degree
  two vertex when considering, exclusively, the sets of edges $\{\e{v_1}{v_2},
  \e{v_3}{v_2}\}$, $\{\e{v_4}{v_2}, \e{v_5}{v_2}\}$, or $\{\e{v_6}{v_2},
  \e{v_7}{v_2}\}$.
  These three sets of edges highlight situations in which $v_2$ can be
  witnessed in different ways.}
  \label{fig:deg2scenarios}
\end{figure}

\begin{proposition}[Same Quadrant]
If $v_1$ and $v_3$ lie in the same quadrant, such as the vertices
$v_4$ and $v_5$ in~\figref{deg2scenarios},
then~$v_2$ will be witnessed in ECCs from every one of the cardinal directions.
\label{prop:deg2SameQuad}
\end{proposition}

\begin{proposition}[Neighboring Quadrants]
If $v_1$ and $v_3$ lie in neighboring quadrants, such as vertices
$v_1$ and $v_3$ in~\figref{deg2scenarios},
then $v_2$ will be witnessed in ECCs from exactly two of the four cardinal directions.
\label{prop:deg2NeighborQuad}
\end{proposition}
\begin{proposition}[Degree Two Bounded Angle]
If $angle(\e{v_1}{v_2}, \e{v_2}{v_3}) < \frac{\pi}{2}$ then $v_2$ will
be witnessed in ECCs from at least two of the four cardinal directions.
\label{prop:deg2Angle}
\end{proposition}
The above propositions show scenarios for which degree two vertices
can be witnessed using cardinal directions.
However, degree two vertices pose particular problems
when the edges lie in non-neighboring quadrants, such as the edges
$\e{v_6}{v_2}$ and $\e{v_7}{v_2}$ in \figref{deg2scenarios}
or $\e{v_1}{v_2}$ and $\e{v_2}{v_3}$ in \figref{deg2Colin}.
Then, when degree two vertices are not included in a plane graph
$\simComp$, a constant number of ECCs can be used to
determine the embeddings of the vertices.

\section{A Special Case}
\label{sec:noDegTwo}
If a plane graph contains no degree two vertices, the graph can be
reconstructed using a finite number of ECCs.  Let $\simComp$ denote a plane
graph with vertex and edge sets $V$ and $E$ respectively.  Recall from
\cite{belton2018learning} that three way filtration line intersections from
carefully chosen directions correspond to a vertex location for plane graphs
using persistence diagrams.  We show that this result still holds for reconstructing
plane graphs using ECCs, if they do not contain degree two vertices.
The proofs of the following lemmas and theorem
can be found in \appendref{proofs}.

First, we provide a lemma that yields insight into how non-degree two
vertices are witnessed.
\begin{lemma}[Linear Witness Lines]
Let $\simComp$ be a plane graph in $\R^2$ with vertices $V$ such
that for all $v \in V$, $deg(v) \neq 2$ and denote $|V|=n$.
Let $\ell$ be a line in $\R^2$ such that any line parallel to $\ell$
intersects at most one vertex in $V$.
Let $\dir \in \sph^1$ be chosen perpindicular to $\ell$.
Then, $$|\pLines{\dir}{V} \cup \pLines{-\dir}{V}| = n$$
\label{lem:vertFiltLine}
\end{lemma}
By generalizing the results of \lemref{vertFiltLine},
we introduce the the following Lemma to generate
$n^2$ potential vertex locations in $\R^2$, where $n$ is the number of vertices.
\begin{lemma}[Witness Line Intersections]
Recall the cardinal directions $\e{0}{1},\e{1}{0},\e{0}{-1},\e{-1}{0}\in \sph^1$.
If for all $v\in V$, $deg(v)\neq 2$ then
\begin{align*}
|\pLines{\e{0}{1}}{V} \cup \pLines{\e{0}{-1}}{V}| &=n \text{, and} \\
|\pLines{\e{1}{0}}{V} \cup \pLines{\e{-1}{0}}{V}| &=n.
\end{align*}
\label{lem:noDeg2SeenVertices}
\end{lemma}
Utilizing these $n$ horizontal and $n$ vertical witness lines, we
are able to pick two additional directions to generate three-way filtration
line intersections using a technique similar to the one described in Theorem~$5$ of \cite{belton2018learning}.
Then, the following theorem holds as well.
\begin{theorem}[ECC Vertex Reconstruction]
Let $\simComp=\langle V,E\rangle$ be a plane graph with vertices $V$ and
edges $E$.
If for all $v\in V$, $deg(v)\neq 2$ then the locations of all vertices can be
determined using six ECCs in $O(n\log n)$ time.
\label{thm:noDeg2VertexReconstruct}
\end{theorem}
The proof of \thmref{noDeg2VertexReconstruct} is found in \appendref{proofs}, but note that
the result follows using similar arguments to those found in Theorem $5$ of
\cite{belton2018learning}.

\section{Discussion and Future Work}
\label{sec:disc}

We have shown that, for any known plane graph $\simComp$, we can choose a linear
number of directions to fully describe $\simComp$ using only ECCs from those
directions. However, when $\simComp$ is unknown, determining such a set is difficult.
We emphasize that although there is an infinite number of directions
in which the vertices of a plane graph can be witnessed by an ECC, the presence
of degree two vertices can restrict these directions to an arbitrarily small
subset of $\sph^1$.

Our ultimate goal is to further develop
the theory on determining the minimal set of
directions necessary to reconstruct shapes.
We are currently investigating upper
bounds on the number of directions needed to reconstruct a plane graph from
ECCs. Additionally, we are exploring what
assumptions we can place on the underlying shape in order to overcome the
challenges of degree two vertices. For example, we observe that if the number of
vertices $|V| = n$ is known, then the intersection of $m> n$ filtration lines
determines the location of all vertices.
Another simplifying assumption is that
minimum angle between any three vertices, $\epsilon$, is known. Then, we can
avoid some of the issues described in \secref{degTwoProb}~by employing pairs of directions
whose difference in angle is less than $\epsilon$.  Finally, we would like to
extend our work to more general shapes embedded in
$\R^d$.


\paragraph*{Acknowledgements}
This material is based upon work supported by the National Science Foundation
under Grant No.\ CCF 1618605 (authors BTF and SM) and Grant No.\ DBI 1661530;
BTF and AS acknowledge the support of NIH and NSF under Grant No.\ NSF-DMS 1664858.

\bibliographystyle{plain}
\bibliography{references}

\begin{thebibliography}{1}

\bibitem{belton2018learning}
Robin~Lynne Belton, Brittany~Terese Fasy, Rostik Mertz, Samuel Micka, David~L
  Millman, Daniel Salinas, Anna Schenfisch, Jordan Schupbach, and Lucia
  Williams.
\newblock Learning simplicial complexes from persistence diagrams.
\newblock {\em arXiv preprint arXiv:1805.10716}, 2018.

\bibitem{crawford2016functional}
Lorin Crawford, Anthea Monod, Andrew~X. Chen, Sayan Mukherjee, and Raúl
  Rabadán.
\newblock Functional data analysis using a topological summary statistic: {T}he
  smooth {E}uler characteristic transform.
\newblock arXiv:1611.06818, 2016.

\bibitem{curry2018directions}
Justin Curry, Sayan Mukherjee, and Katharine Turner.
\newblock How many directions determine a shape and other sufficiency results
  for two topological transforms.
\newblock arXiv:1805.09782, 2018.

\bibitem{edelsbrunner2010computational}
Herbert Edelsbrunner and John Harer.
\newblock {\em Computational Topology: {A}n Introduction}.
\newblock American Mathematical Society, 2010.

\bibitem{ghrist2018euler}
Robert Ghrist, Rachel Levanger, and Huy Mai.
\newblock Persistent homology and {E}uler integral transforms.
\newblock arXiv:1804.04740, 2018.

\bibitem{Tangelder2007}
Johan W.~H. Tangelder and Remco~C. Veltkamp.
\newblock A survey of content based 3d shape retrieval methods.
\newblock {\em Multimedia Tools and Applications}, 39(3):441, Dec 2007.

\bibitem{turner2014persistent}
Katharine Turner, Sayan Mukherjee, and Doug~M. Boyer.
\newblock Persistent homology transform for modeling shapes and surfaces.
\newblock {\em Information and Inference: A Journal of the IMA}, 3(4):310--344,
  2014.

\end{thebibliography}

\appendix
\section{Proofs}\label{append:proofs}

\paragraph{Proof of \propref{nDirExist}}(ECC Existence)
\begin{proof}

Let $v\in V$ be a vertex in $\simComp$.  First, we show that each vertex is
witnessed from an infinite number of directions $\sph^1$.  If $deg(v) =
0$, $v$ is witnessed from any direction for which it lies on a unique
    witness line (So, for all but $|V|-1$ directions). If $deg(v) = 1$ with edge $\e{v}{v'}$ for some $v'\in V$,
then $v$ is observed from an the infinite set of directions from which
$v'$ appears after $v$ in the lower-star filtration, and $v$ lies on a unique witness line. If
$deg(v) > 1$ with edges $\e{v}{v'}$ and $\e{v}{v''}$ for $v',v''\in
V$, then~$v$ is observed from any direction from which $v'$ and~$v''$
appear before $v$ in the filtration and $v$ lies on a unique witness
line.  Thus, each vertex is witnessed from an infinite number of
directions.

Let $\mathbb{I}_v$ be the set of directions that witness $v$. We can
choose any three directions from $\mathbb{I}_v$ and generate a unique
three-way intersection at $v$.
Now, we need to show that a set of directions exist for each of the $n$ vertices
such that no three-way intersections exist at locations
where a vertex is not located. In order to do this, we give the vertices some
arbitrary ordering $v_1, v_2, \ldots, v_n$. Then, select vertices in ascending
order. For the first, any three directions in $\mathbb{I}_{v_1}$ will give
a single three-way intersection of witness lines. For each successive
vertex $v_i$, there exist up to $3i^2$ witness lines.
More importantly, the number of three-way witness line
intersections is finite. Thus, there exist three directions in
$\mathbb{I}_{v_i}$ such that none of the witness lines created by these
directions intersect existing intersections.
Since the $x$- and $y$-coordinates of a vertex can be determined
using a three-way line intersection, we can see that there
exist a set of $3n$ directions which generates exactly $n$ three-way
intersections of witness lines, revealing the location of all $n$
vertices.

\end{proof}

\paragraph{Proof of \propref{deg2SameQuad}} (Same Quadrant)
\begin{proof}
If $v_1$ and $v_3$ lie in the same quadrant, then
$v_1$ and $v_3$ will appear before $v_2$ from exactly
one of the two $x$-axis parallel directions directions $\e{-1}{0}$ or $\e{1}{0}$
and before $v_2$ in exactly one of the $y$-axis parallel directions
$\e{0}{-1}$ or $\e{0}{1}$.
Let $\dir_1 \in \{\e{0}{1},\e{0}{-1}\}$ and
$\dir_2 \in \{\e{1}{0},\e{-1}{0}\}$ be the directions
that witness $v_1$ and $v_3$ before $v_2$.
$\ecc{\dir_1}{\simComp}$ and $\ecc{\dir_2}{\simComp}$ will witness $v_2$
by seeing a decrease in the Euler Characteristic at the
time that $v_2$ is first included in the filtration.
Then, $-\dir_1$ and $-\dir_2$ will witness $v_2$ before
$v_1$ or $v_2$.
Since no other edges with $v_2$ as an endpoint exist, there
will be an increase in $\ecc{-\dir_1}{\simComp}$ and $\ecc{-\dir_2}{\simComp}$
at the time that $v_2$ is first included in the filtration.
Then, $v_2$ is witnessed from every cardinal direction, as required.
\end{proof}

\paragraph{Proof of \propref{deg2NeighborQuad}} (Neighboring Quadrants)
\begin{proof}
	Recall that no two vertices share $x$- or $y$-coordinates, then
	any witness line from a cardinal direction will be unique.
    Let $s$ be the cardinal direction for which $v_1\cdot s < v_2 \cdot s$
	and $v_3\cdot s > v_2 \cdot s$
	and $-s$ the cardinal direction chosen such that $v_3\cdot s < v_2 \cdot s$
	and $v_1\cdot s > v_2 \cdot s$.
	Then, there is no change
	in Euler Characteristic at $v_2$ from either $s$ or $-s$, since $v_2$ is
	added at the same time as $(v_1,v_2)$ or $(v_2,v_3)$, respectively. Now,
	let $w$ and $-w$ be the remaining two cardinal directions, where $w$ is
	the direction from which we include $v_2$ before $v_1$ or $v_3$. Direction $w$
	witnesses $v_2$ because no edges are included at height $v_2$ from
	that direction. Direction $-w$ witnesses $v_2$ because both $(v_1,v_2)$ and
	$(v_2,v_3)$ are added along with $v_2$. Thus, $v_2$ is witnessed from
	exactly two of the four cardinal directions.
\end{proof}

\paragraph{Proof of \propref{deg2Angle}} (Degree Two Bounded Angle)
\begin{proof}
If $angle(\e{v_1}{v_2}, \e{v_2}{v_3}) < \frac{\pi}{2}$, then $v_1$
and $v_3$ must lie in neighboring quadrants or the same quadrant, since neither
can lie on the boundary of a quadrant by assumption. If they are in
the same quadrant, \propref{deg2SameQuad} tells us that they must be
seen from all four cardinal directions. If they are in neighboring
quadrants, \propref{deg2NeighborQuad} tells us that we can witness
$v_2$ with ECCs from exactly two of the four cardinal directions.
\end{proof}

\paragraph{Proof of \lemref{vertFiltLine}} (Linear Witness Lines)
\begin{proof}
We show that each vertex is seen by at least one of $\dir$ or $-\dir$.
Let $v\in V$ be a vertex with $deg(v)=0$.
Then, $v$ will correspond to $\pLine{\dir}{v}$ for any
arbitrary direction $\dir \in \sph^1$
because $\ecc{\dir}{\simComp}$ will always increase by at least one
at time $\dir \cdot v$.
As such, $v$ will be observed by both $\dir$ and $-\dir$.

Let $v\in V$ be a vertex with $deg(v)=1$ and $\e{v}{v'}\in E$ for
some $v'\in V$.
Then, if $\dir \in \sph^1$ is chosen such that $\dir \cdot v' < \dir \cdot v$,
$v$ will not result in a change in $\ecc{\dir}{\simComp}$.
However, $\dir$ was chosen such that no two vertices will be
observed at the same time.
As a result, no edge in $E$ can be parallel to $\ell$.
Then,
if $\dir \cdot v' < \dir \cdot v$ then $-\dir \cdot v' \ge -\dir \cdot v$
and an increase in $\ecc{-\dir}{\simComp}$ is seen at time $-\dir \cdot v$.
This implies that $v$ is observed by $\dir$ or $-\dir$ but not both.

Finally, if $v\in V$ is a vertex with $deg(v) > 2$, then we must consider
two cases.
If, for $\dir \in \sph^1$,
there exists exactly one edge $\e{v}{v'}\in
E$ such that $\dir \cdot v' < \dir \cdot v$, then there must
exist at least two additional edges that will result in a decrease
in $\ecc{-\dir}{\simComp}$ at time $-\dir \cdot v$.
As such, $v$ will be observed by at least one of the ECCs resulting
from $\dir$ or $-\dir$.
On the other hand, if, for $\dir \in \sph^1$,
there exists either zero edges or more than one edge
that appear before $v$ in the height filtration from $\dir$,
then $\ecc{\dir}{\simComp}$ will either increase (in the case where
no edges appear before $v$) or decrease (in the case where two
or more edges appear before $v$).
Then, all non-degree two vertices result in a change in $\ecc{\dir}{\simComp}$
or $\ecc{-\dir}{\simComp}$ and, as such, $|\pLines{\dir}{V} \cup \pLines{-\dir}{V}|
= n$, as required.
\end{proof}

\paragraph{Proof of \lemref{noDeg2SeenVertices}} (Witness Line Intersections)
\begin{proof}
By \lemref{vertFiltLine}, if $\dir$ is chosen such that
no two vertices are intersected by a line perpendicular to $\dir$,
then $\pLines{\dir}{V}\cup \pLines{-\dir}{V}$ will result in
$n$ filtration lines.
Recall that no two vertices in $\simComp$ share an $x$- or $y$-coordinate.
Then, by \lemref{vertFiltLine},
$|\pLines{\e{0}{1}}{V} \cup \pLines{\e{0}{-1}}{V}| =n$ and
$|\pLines{\e{1}{0}}{V} \cup \pLines{\e{-1}{0}}{V}| =n$, as required.
\end{proof}

\paragraph{Proof of \thmref{noDeg2VertexReconstruct}} (ECC Vertex Reconstruction)
\begin{proof}
Using \lemref{noDeg2SeenVertices} we construct $n$ horizontal and $n$ vertical
lines corresponding to vertices using four ECCs and we denote them $L_H$ and
$L_V$ respectively.
Then, we must identify an additional two directions which will, together,
generate an additional $n$ unique witness lines and exactly $n$ three-way filtration
line intersections.
We choose these final directions $\dir_3 \in \sph^1$ and
$-\dir_3$ using the method described in
Theorem 5 of \cite{belton2018learning}.
We observe that, by Lemma 4 of \cite{belton2018learning},
no two vertices will be intersected
by any single line perpindicular to $\dir_3$.
Then, since each vertex will be witnessed by at least one of the ECCs from
$\pLines{\dir_3}{V}$ or $\pLines{-\dir_3}{V}$ by \lemref{vertFiltLine},
these two directions will yield
$n$ distinct filtration lines each of which will intersect exactly one
two-way intersection between lines of $L_H$ and $L_V$.
Then, Lemma 3 of \cite{belton2018learning} implies that these three-way intersections are
the locations of the $n$ vertices in $V$.
The $O(n\log n)$ running time follows from the proof of Theorem 5 in
\cite{belton2018learning}.
\end{proof}

\end{document}